# Electrical and Thermal Transport Properties of the β-Pyrochlore Oxide $CsW_2O_6$

Yoshihiko Okamoto*, Kenta Niki, Rikuto Mitoka, and Koshi Takenaka

*Department of Applied Physics, Nagoya University, Nagoya 464-8603, Japan*

We report the electrical resistivity, thermoelectric power, and thermal conductivity of single-crystalline and sintered samples of the $5d$ pyrochlore oxide $CsW_2O_6$. The electrical resistivity of the single crystal is 3 mΩ cm at 295 K and gradually increases with decreasing temperature above 215 K (Phase I). The thermoelectric power of the single-crystalline and sintered samples shows a constant value of approximately −60 μV K$^{-1}$ in Phase I. These results reflect that the electron conduction by W $5d$ electrons in Phase I is incoherent and in the hopping regime, although a band gap does not open at the Fermi level. The thermal conductivity in Phase I of both samples is considerably low, which might be due to the rattling of Cs$^+$ ions. In Phase II below 215 K, the electrical resistivity and the absolute value of thermoelectric power of both samples strongly increase with decreasing temperature, corresponding to a transition to a semiconducting state with a band gap open at the Fermi level, while the thermal conductivity in Phase II is smaller than that in Phase I.

## I. INTRODUCTION

Pyrochlore oxides with $5d$ transition metal elements are known to show interesting electronic properties unique to the $5d$ electrons, such as a metal-insulator transition accompanied by all-in-all-out type magnetic order in $Cd_2Os_2O_7$ and $Nd_2Ir_2O_7$ [1-3], a parity violating transition in $Cd_2Re_2O_7$ [4,5], and rattling superconductivity in $KOs_2O_6$ [6,7]. The effects of strong electron correlation and strong spin-orbit coupling in $5d$ electron systems appear in various physical properties through the pyrochlore structure made of regular tetrahedra. $CsW_2O_6$ is a $5d$ pyrochlore oxide with the β-pyrochlore structure. It is therefore important to elucidate its physical properties. $CsW_2O_6$ was first synthesized by Cava et al. and was reported to have a cubic lattice with the $Fd\text{–}3m$ space group at room temperature [8]. In this crystal structure, as shown in Fig. 1(a), W atoms are octahedrally coordinated by oxygen atoms and the $WO_6$ octahedra are connected via corner sharing. As shown in Fig. 1(b), the pyrochlore structure is formed when the nearest-neighbor W atoms are connected to each other. The W atoms have a valence of 5.5+ with a $5d^{0.5}$ electron configuration.

Recently, $CsW_2O_6$ was reported to show an electronic phase transition accompanied by a structural change from cubic $Fd\text{–}3m$ to cubic $P2_13$ at $T_t$ = 215 K by structural refinement and physical property measurements on single crystals [9]. The phases above and below $T_t$ were named Phases I and II, respectively. The space group of Phase II was first reported to be orthorhombic $Pnma$ based on powder diffraction data [10], but a theoretical study [11] suggested that this is incorrect. In Phase II, $W_3$ trimers with a regular triangle shape are formed in the pyrochlore structure. This trimer formation represents a complex self-organization of $5d$ electrons, which can be resolved into a charge order satisfying the Anderson condition in a nontrivial way, orbital order of $5d$ orbitals, and the formation of a spin-singlet pair in a regular-triangle trimer. As a result, Phase II was reported to be a nonmagnetic insulator. The phase transition between Phases I and II was also observed in a study on thin films [12]. Although the electrical resistivity ρ of Phase I is much smaller than that of Phase II in all the measured samples, the ρ in Phase I increases with decreasing temperature [9,10,12]. This corresponds to the optical conductivity data of Phase I, which indicates that the coherency of the $5d$ electrons is almost lost [9]. The strong correlation for a $5d$ electron system likely plays an important role in the nonmetallic behaviors in Phase I.

$CsW_2O_6$ has the potential to show high thermoelectric performance but there have been no reports on its thermoelectric power $S$ and thermal conductivity κ. When a material simultaneously has large $S$ and low ρ and κ, it is considered a promising thermoelectric material. Practical materials, such as $Bi_2Te_3$-based materials, show a dimensionless figure of merit $ZT = S^2T/\rho\kappa$ of ~1 [13], and new materials that exhibit larger $ZT$ are desired. $CsW_2O_6$ can exhibit large $|S|$ while keeping low ρ, as in the cases of Co, Rh, and Ti oxides [14-22], because the $t_{2g}$ band in $CsW_2O_6$ is occupied by a small number of $5d$ electrons. Another reason for the potential high thermoelectric performance is a rattling effect, which is known as a mechanism that significantly reduces the lattice thermal conductivity in clathrate compounds and filled skutterudites [23-26]. Cs$^+$ ions in $CsW_2O_6$ are located in a highly-symmetric and oversized $O_{18}$ cage, as in the cases of $K^+$ and $Rb^+$ ions in $KOs_2O_6$ and $RbOs_2O_6$, respectively [6,7,27], suggesting that the rattling effect can be significant in $CsW_2O_6$.

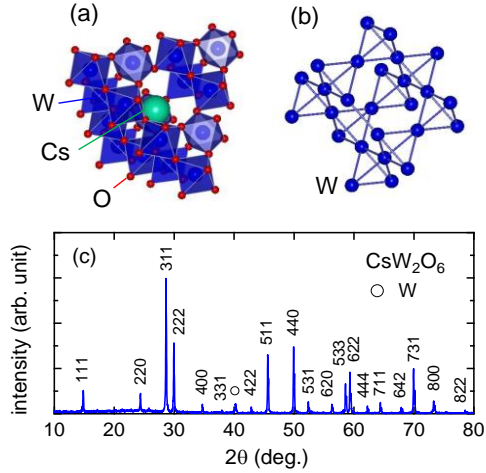

Fig. 1. (a) Crystal structure of $CsW_2O_6$. (b) Pyrochlore structure of W atoms in $CsW_2O_6$. (c) A powder X-ray diffraction pattern of a $CsW_2O_6$ powder sample taken at room temperature, indexed by a cubic cell of $a = 10.3215(5)$ Å. The peak indicated by an open circle is a diffraction peak from a W impurity.

In this study, we report the ρ, $S$, and κ of single-crystalline and sintered samples of $CsW_2O_6$. These measurements are not only important for evaluating the performance as a thermoelectric material but are also essential to elucidate the physics behind the electronic phase transition in $CsW_2O_6$. The maximum $ZT$ value of 0.014 was observed in a single crystal at 290 K. In Phase I, the ρ gradually increases with decreasing temperature and the $S$ shows a constant value of approximately −60 μV K$^{−1}$, suggesting that the hopping electron conduction due to strong electron correlation for a 5$d$ electron system is realized. The single-crystalline and sintered samples showed small κ, likely due to the effect of rattling. The lattice thermal conductivity of the single crystal is estimated to be 15 mW cm$^{−1}$ K$^{−1}$ at 290 K and becomes smaller in Phase II.

## II. EXPERIMENTS

Single crystals of $CsW_2O_6$ were prepared by crystal growth in an evacuated quartz tube under a temperature gradient [9]. A mixture of a 3:1:3 molar ratio of $Cs_2WO_4$ (Alfa Aeser, 99.9%), $WO_3$ (Kojundo Chemical Laboratory, 99.99%), and $WO_2$ (Kojundo Chemical Laboratory, 99.99%), with a combined mass of 0.1 g, was sealed in an evacuated quartz tube with 0.1 g of CsCl (Wako Pure Chemical Corporation, 99.9%). The hot and cold sides of the tube were heated to and kept at 973 and 873 K for 96 h, respectively, and then the furnace was cooled to room temperature. The mixture was put on the hot side. The obtained single crystals had an octahedral shape with edges of at most several mm. Powder samples of $CsW_2O_6$ were prepared by a solid-state reaction method [8,10]. A mixture of a 3:1:3 molar ratio of $Cs_2WO_4$, $WO_3$, and $WO_2$ was sealed in an evacuated quartz tube. The tube was quickly heated to and kept at 873 K for 24 h, and then quenched to room temperature. The obtained sample contains $CsW_2O_6$ and water-soluble $Cs_2WO_4$ phases, the latter of which was removed by rinsing the sample with deionized water. The obtained $CsW_2O_6$ powder was sintered at 773 K for 10 min using a spark plasma sintering furnace (SPS Syntex). Sample characterization was performed by powder X-ray diffraction analysis with Cu Kα radiation at room temperature using a RINT-2100 diffractometer (RIGAKU). As shown in Fig. 1(c), all diffraction peaks, except for a small peak due to a W impurity, were indexed on the basis of a cubic cell of $a = 10.3215(5)$ Å, indicating that β-pyrochlore-type $CsW_2O_6$ was obtained as an almost single phase. The electrical resistivity of single-crystalline and sintered samples of $CsW_2O_6$ were measured by a four-probe method. The thermoelectric power and thermal conductivity of the single-crystalline and sintered sample were measured by a steady-state method. Some of the thermoelectric power measurements were performed using a Physical Property Measurement System (Quantum Design). Crystal structure views were drawn using VESTA [28].

## III. RESULTS

Figure 2(a) shows the temperature dependence of ρ of the single-crystalline and sintered samples of $CsW_2O_6$. For both samples, ρ gradually increases with decreasing temperature from room temperature. The temperature derivative of ρ, $d\rho/dT$, is negative in sintered samples, but also in single crystals, suggesting that the observed temperature dependence is intrinsic. At lower temperatures, ρ rapidly increases below $T_t = 215$ K. As seen in Fig. 2(a), ρ exponentially increases with decreasing temperature below ~200 K. As shown in the Arrhenius plot of the electrical conductivity (σ = 1/ρ) shown in the inset of Fig. 2(a), linear behavior appears in a low temperature region in both samples. The activation energies are estimated to be $E_a$ = 94.9(2) and 93.2(2) meV for the single-crystalline and sintered sample, respectively. The almost identical $E_a$ values in both samples suggest that the obtained $E_a$ is intrinsic and that Phase II is a semiconductor with a band gap corresponding to this $E_a$ value. The ρ of the sintered sample is 66 mΩ cm at 295 K, which is more than twenty times larger than that of the single crystal of ρ = 3 mΩ cm, meaning that the influence of grain boundaries on electrical conduction is considerably strong. However, the ρ value of the sintered sample in this study is almost 3% of that of the previous study, where spark plasma sintering was not performed [10], indicating that this method is effective for improving the sinterability of this material.

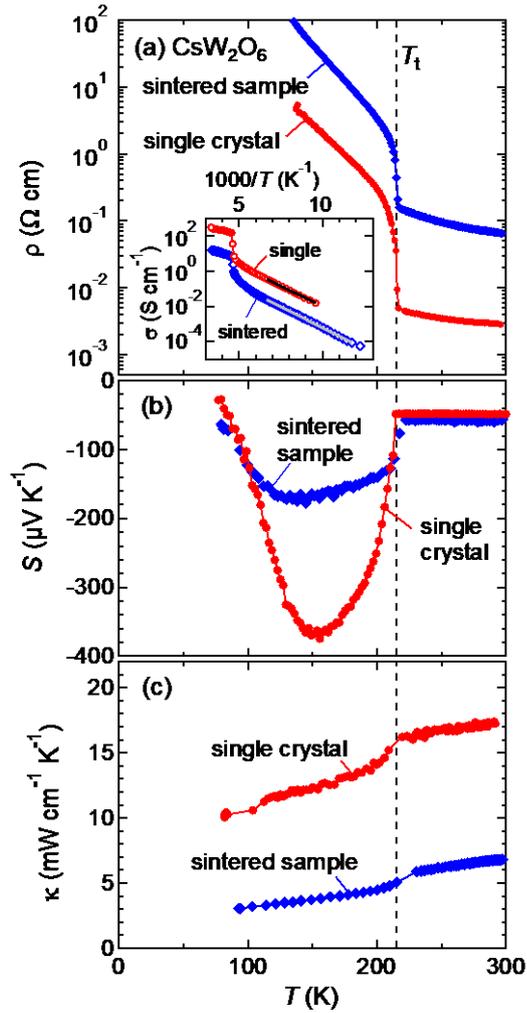

Fig. 2. Temperature dependences of electrical resistivity (a), thermoelectric power (b), and thermal conductivity (c) of single-crystalline and sintered samples of $CsW_2O_6$. The broken lines indicate the electronic transition temperature, $T_t = 215$ K. The inset in (a) shows Arrhenius plots of the electrical conductivity of single-crystalline and sintered sample. Solid lines represent the results of linear fits to 104–150 and 85–150 K data for single-crystalline and sintered samples, respectively.

Figure 2(b) shows the temperature dependence of $S$ of $CsW_2O_6$ single-crystalline and sintered samples. In Phase I, $S$ shows a constant value of $-50$ µV K$^{-1}$ for a single crystal and $-60$ µV K$^{-1}$ for a sintered sample. The negative $S$ values indicate that the electron carriers are dominant in Phase I. A previous study on thin films reported the presence of holes in Phase I by Hall measurements [12]. At present, the reason why the results of $S$ and Hall measurements do not simply match is an open question. On the other hand, the temperature independent $S$ in Phase I suggests that it is not metallic. In general, $S$ of metals is proportional to temperature. As discussed below, regarding other physical properties such as electrical resistivity and optical conductivity, the electrical conduction represented by a hopping picture is realized in Phase I, rather than coherent electron conduction.

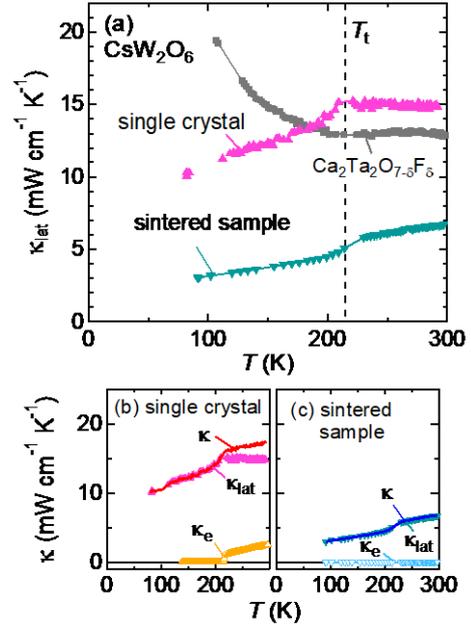

Fig. 3. (a) Temperature dependence of lattice thermal conductivity of single-crystalline and sintered samples of $CsW_2O_6$. The κ data for $Ca_2Ta_2O_{7-\delta}F_\delta$ sintered sample are also shown [37]. The broken line indicates $T_t = 215$ K. (b, c) Estimation of lattice thermal conductivity of single-crystalline and sintered samples of $CsW_2O_6$. Each panel shows the measured κ, electron thermal conductivity $\kappa_e$ estimated by applying the Wiedemann–Franz law to the electrical resistivity data, and lattice thermal conductivity $\kappa_{lat} = \kappa - \kappa_e$.

As shown in Fig. 2(b), $S$ of Phase II shows large and negative values. Below $T_t$, $|S|$ rapidly increases with decreasing temperature and exhibits maximum values of 370 and 170 µV K$^{-1}$ for single-crystalline and sintered samples, respectively. This behavior corresponds to the semiconducting nature of Phase II with a reduced carrier density compared to Phase I. Unlike Phase I, there is a considerable sample dependence in the $S$ values of Phase II. It is natural to consider that Phase II is an electron-doped semiconductor, where electron carriers are introduced by some lattice defects, and the number of defects differs depending on the sample. In a previous study of $CsW_2O_6$, it was reported that W deficiency occurred for some synthesis conditions [9], but it simply results in hole doping. Therefore, negative $S$ values with strong sample dependence in Phase II are expected to be due to another factor. Since it is difficult for Cs and W atoms to be present in excess in the case of the β-pyrochlore structure, the most likely scenario is oxygen deficiency. To the contrary, there is no strong sample dependence in the $S$ values of Phase I, suggesting that the band gap does not open at $E_F$ in Phase I.

The temperature dependence of κ of the $CsW_2O_6$ single-crystalline and sintered samples is shown in Fig. 2(c). The values of κ are considerably small: 17 mW cm$^{-1}$ K$^{-1}$ for a single crystal at 290 K and 7 mW cm$^{-1}$ K$^{-1}$ for a sintered sample at 295 K. For both samples, $\kappa$ further decreases with decreasing temperature. Figure 3 shows the lattice thermal conductivity $\kappa_{lat}$ of both samples obtained by subtracting the electron contribution from the measured $\kappa$, assuming that the electron and phonon contributions are independent and the total thermal conductivity is the summation of these contributions. Details of the subtraction are shown in Figs. 3(b) and (c). The electron thermal conductivity $\kappa_e$ was estimated by applying the Wiedemann–Franz law, $\kappa_e\rho = LT$, where $L$ is the Lorentz number of $2.44 \times 10^{-2}$ mW m$\Omega$ K$^{-2}$, to the $\rho$ data. The estimated $\kappa_{lat}$ of both samples are considerably low and comparable to those of Ge clathrate compounds, where $\kappa_{lat}$ is strongly suppressed by the rattling effect [23,24].

## IV. DISCUSSION
### 4.1 Dimensionless figure of merit ZT

We discuss the dimensionless figure of merit $ZT = S^2T/\rho\kappa$ of CsW$_2$O$_6$. Both the single-crystalline and sintered samples showed the largest $ZT$ values at the highest measured temperatures. The maximum $ZT$ values of the single-crystalline and sintered samples are 0.014 at 290 K and 0.0024 at 295 K, respectively. These values are much smaller than the practical level of $ZT = 1$. As shown in Fig. 2(c), $\kappa$ of CsW$_2$O$_6$ is low enough as a thermoelectric material, yet the $ZT$ of CsW$_2$O$_6$ remains small, mainly due to small $|S|$. In addition, $\rho$ of the sintered sample is too large. The practical materials show $\rho$ of 1–2 m$\Omega$ cm [13]. The value of $\rho = 3$ m$\Omega$ cm for the single crystal at 295 K is sufficiently low as a thermoelectric material and the $\rho$ will be lower at a higher temperature, but $\rho = 66$ m$\Omega$ cm for the sintered sample is too large. In order to use a CsW$_2$O$_6$ sintered sample as a thermoelectric material, it is essential to reduce $\rho$ by improving the sinterability.

### 4.2 Thermoelectric power in Phase I

The single-crystalline and sintered samples of CsW$_2$O$_6$ showed almost temperature independent $S$ of about $-60$ μV K$^{-1}$ in Phase I. This temperature independent $S$ is in contrast to those of mixed-valent spinel compounds, such as CuIr$_2$S$_4$, LiRh$_2$O$_4$, and CuTi$_2$S$_4$, which have a pyrochlore structure made of transition metal atoms with a half-integer valence, as in the case of CsW$_2$O$_6$. The former two compounds have an almost fully occupied $t_{2g}$ band and show a metal-insulator transition accompanied by charge order. Above the transition temperature, they show positive $S$ increasing with increasing temperature [29,30]. CuTi$_2$S$_4$ has a $t_{2g}^{0.5}$ electron configuration, the same as for CsW$_2$O$_6$, and shows small and negative $S \sim -10$ μV K$^{-1}$ at 290 K [31,32]. The value of $S$ decreases with increasing temperature and reaches $-20$ μV K$^{-1}$ at 600 K. The temperature independent $S$ and $d\rho/dT < 0$ in Phase I of CsW$_2$O$_6$ mean that the electrical conduction in this phase is represented by a hopping picture, where the coherency of conduction electrons is lost. This is more directly evidenced by the optical conductivity of Phase I, which is characterized by a large infrared absorption with a broad peak at $\sim$0.6 eV and an indistinct Drude contribution [9]. The conducting electrons are trapped by something with an energy scale of 0.6 eV, resulting in the loss of coherency. This behavior is likely caused by the strong electron correlation due to the small orbital overlap in CsW$_2$O$_6$, where WO$_6$ octahedra are connected by corner sharing with a bent W-O-W bond. This situation is different from the spinel structure made of edge-shared octahedra.

When the electrical conduction is in the hopping regime, $S$ shows a constant value given by the Heikes formula. The measured $S$ in Phase I of CsW$_2$O$_6$ coincides with $S_H = -(k_B/e)$ ln $[2(1 - x)/x] = -60$ μV K$^{-1}$, which is the generalized Heikes formula for the case where the spin degrees of freedom of the conduction electrons are considered but the double occupancy of an orbital is forbidden by strong on-site Coulomb repulsion [33]. In this formula, $x$ is the ratio of occupied W sites and equals 0.5 for CsW$_2$O$_6$. This result supports the fact that in Phase I of CsW$_2$O$_6$, 5$d$ electrons have a localized nature due to strong electron correlation for 5$d$ electron systems, and the electrical conduction is in the hopping regime. The temperature independent behavior of the order of $S \sim -60$ μV K$^{-1}$ was observed in tetracyanoquinodimethane (TCNQ) salts with a half-integer valence [34,35]. In contrast, the origin of the large $S$ in Co oxides was explained based on electron hopping with the spin and orbital degrees of freedom of $t_{2g}$ orbitals, which gives large $S_H = (k_B/e)$ ln $[6x/(1 - x)]$ [36], but such a situation is not realized in CsW$_2$O$_6$. However, the fact that $S$ is represented by the generalized Heikes formula means that it can be significantly enhanced by controlling the electron filling, which might improve the thermoelectric performance of this compound. For example, if $x$ is decreased to be 0.15 by hole doping, $S$ is expected to reach $-210$ μV K$^{-1}$, comparable to the $S$ of practical materials.

### 4.3 Lattice thermal conductivity

As shown in Fig. 3(a), single-crystalline and sintered samples of CsW$_2$O$_6$ showed low $\kappa_{lat}$ of 15 and 7 mW cm$^{-1}$ K$^{-1}$ at around room temperature, respectively, with unusual temperature dependence, different from the case of the normal crystalline solids. The $\kappa_{lat}$ of the sintered sample is most likely suppressed by the grain boundaries. However, the $\kappa_{lat}$ of both single-crystalline and sintered samples is also suppressed by intrinsic effects for this material, which are highlighted by comparing the results of CsW$_2$O$_6$ and α-pyrochlore oxide Ca$_2$Ta$_2$O$_{7-\delta}$F$_\delta$ [37]. Ca$_2$Ta$_2$O$_{7-\delta}$F$_\delta$ has similar elements and crystal structure to CsW$_2$O$_6$. The measured $\kappa$ of Ca$_2$Ta$_2$O$_{7-\delta}$F$_\delta$ is equal to $\kappa_{lat}$ because it is an insulator. However, rattling does not occur in this compound, because the large Ca$_4$ tetrahedron instead of the $Cs^+$ ion is accommodated in the $O_{18}$ cage. As shown in Fig. 3(a), the $\kappa$ of a $Ca_2Ta_2O_{7-\delta}F_\delta$ sintered sample is 13 mW cm$^{-1}$ K$^{-1}$ at 300 K, which is about double the $\kappa_{lat}$ of the $CsW_2O_6$ sintered sample and comparable to that of the $CsW_2O_6$ single crystal. This result suggests that the $\kappa_{lat}$ of $CsW_2O_6$ is reduced by the effect of rattling. In addition, unlike the case of $CsW_2O_6$, the $\kappa$ of $Ca_2Ta_2O_{7-\delta}F_\delta$ is almost inversely proportional to temperature between 100 and 300 K. This behavior reflects that $Ca_2Ta_2O_{7-\delta}F_\delta$ is a normal crystalline solid without rattling, and implies that the unusual temperature dependence of $\kappa_{lat}$ in $CsW_2O_6$ is caused by rattling.

Another possibility that causes the difference in the $\kappa_{lat}$ of $CsW_2O_6$ and $Ca_2Ta_2O_{7-\delta}F_\delta$ is the charge and/or orbital fluctuation. It is known that in materials with a charge and/or orbital order transition, $\kappa_{lat}$ is suppressed by the strong charge and/or orbital fluctuations and shows $d\kappa_{lat}/dT > 0$ above the transition temperature [38-42]. $CsW_2O_6$ exhibits an electronic phase transition with charge and orbital order at 215 K, but $Ca_2Ta_2O_{7-\delta}F_\delta$ does not show such a transition. This difference might appear as a difference in $\kappa_{lat}$ between them. However, it should be noted that the suppression of $\kappa_{lat}$ of $CsW_2O_6$ is more pronounced in Phase II. Low-energy phonons due to the rattling oscillations of the $Cs^+$ ions were observed in the heat capacity data at low temperatures, indicating that the suppression of $\kappa_{lat}$ by the rattling can survive in Phase II [9,10]. In contrast, in the case of suppression by the charge and/or orbital fluctuation, the fluctuation is expected to disappear below the ordering temperature, so $\kappa_{lat}$ must increase below 215 K. This suggests that the suppression of $\kappa_{lat}$ in $CsW_2O_6$ is not due to such a fluctuation. However, large atomic displacement parameters of the O atoms bridging the W atoms in a $W_3$ trimer in Phase II suggests the presence of a strong fluctuation of dimer formation in a $W_3$ trimer [9], which may possibly suppress the $\kappa_{lat}$ in Phase II. There might also be contributions from symmetry lowering and domain formation at the phase transition between Phases I and II. Although the reason why $\kappa_{lat}$ in Phase II is smaller than that in Phase I is still unclear, the very low thermal conductivity in Phase II may lead to a novel method for suppressing the thermal conductivity, especially for thermoelectric materials for low-temperature applications, so the suppression mechanism should be elucidated in future studies.

## V. CONCLUSIONS

We studied the electrical resistivity, thermoelectric power, and thermal conductivity of single-crystalline and sintered samples of the 5$d$ pyrochlore oxide $CsW_2O_6$, which shows an electronic phase transition accompanied by the formation of unique $W_3$ trimers at $T_t$ = 215 K. In Phase II below $T_t$, both samples showed thermally activated electrical resistivity and large and negative thermoelectric power, indicating a semiconducting state with a band gap open at the Fermi level. In contrast, in Phase I above $T_t$, characteristic nonmetallic behaviors were observed, although electrical resistivity is much lower than that in Phase II. In both single-crystalline and sintered samples, the electrical resistivity gradually increases with decreasing temperature. The thermoelectric power shows a constant value of about −60 μV K$^{-1}$, which coincides with the theoretical value given by the generalized Heikes formula in the strongly correlated case. These results suggest that the hopping electron conduction due to strong electron correlation for a 5$d$ electron system is realized in Phase I. The thermal conductivity of $CsW_2O_6$ is considerably small and suppressed by some effects, including rattling. The lattice thermal conductivity of the single crystal is 15 mW cm$^{-1}$ K$^{-1}$ at 290 K and shows smaller values in Phase II.

## ACKNOWLEDGMENTS

The authors are grateful to H. Sawa and N. Katayama for helpful discussions. This work was partly carried out under the Visiting Researcher Program of the Institute for Solid State Physics, the University of Tokyo and supported by JSPS KAKENHI (Grant Numbers: 16H03848, 18H04314, 19H05823, and 20H02603), the Research Foundation for the Electrotechnology of Chubu, and the Asahi Glass Foundation.

*e-mail: yokamoto@nuap.nagoya-u.ac.jp